\documentclass[twocolumn,prb,aps,floats,floatfix]{revtex4}
\usepackage[final]{graphicx}
\usepackage{calc}

\begin{document}
\title{Fractionalization in dimerized graphene and graphene bilayer}
\author{M. V. Milovanovi\'{c}}
\affiliation{Institute of Physics, P.O.Box 68, 11080 Belgrade,
Serbia}
\date{\today}

\begin{abstract}
We show that the fractional statistics of quasiparticles in
dimerized graphene, in recent proposals for charge and statistics
fractionalization, can have two realizations depending whether
elementary objects can be considered as point-like or extended
objects. Therefore, there are two phases of proposed excitations and
we give their topological descriptions with their respective
statistics. We propose that a natural setting for fractionalization
are certain systems with excitonic instabilities and demonstrate
this by an example of graphene bilayer.

\end{abstract}

\maketitle \vskip2pc

\section{Introduction}\label{introduction}
Recent proposals \cite{hou,sar,jackpi,cha} for fractionalization in
dimerized graphene and similar structures are based on the physics
of charge fractionalization in polyacetylene \cite{su}. But
dimerization is hard to achieve in monolayer graphene.
Mathematically fractionalization reduces to finding zero modes in a
Dirac like equation for the dimerized pattern with vortex structure.
A similar problem came from finding zero modes of vortex solutions
of $p$-wave and other superconducting order parameters. From this
comes the idea that in the systems needed for the fractionalization
previously introduced, in which number of electrons is conserved
quantity, excitonic instability with its BCS like hamiltonian and
structure may induce order parameters and vortex solutions with only
one zero mode necessary for fractionalization. Like quantum Hall
bilayer at $\nu = 1$ also graphene bilayer represents a natural
setting for an excitonic instability as found in Ref.
\onlinecite{han}. And similarly but not obviously, as
fractionalization in quantum Hall bilayer into merons with half
electric charge and fractional quantum statistics, we expect similar
quasiparticles - vortex solutions in graphene bilayer.

In Section II we will review the recent proposals for
fractionalization, where we will also clarify the situation and
answer the question pertaining zero mode solutions and statistics
that appeared in literature. Section III will be devoted to the
bilayer graphene as a stage for charge fractionalization.
\section{Fractionalization in dimerized graphene}
The effective (low-energy) Hamiltonian in the presence of the Kekule
deformation of a graphene monolayer is \cite{hou}
\begin{equation}
{\cal H} = \int d^{2} r \Psi^{+}(r) {\cal K}_{D} \Psi(r)\label{ham}
\end{equation}
with
\begin{displaymath}
\Psi^{+}(r) = (u_{b}^{+}(r) u_{a}^{+}(r) v_{a}^{+}(r) v_{b}^{+}(r))
\end{displaymath}
and
\begin{displaymath}
{\cal K}_{D} = \left( \begin{array}{cccc} 0& -2i \partial_{z}&
\Delta(r) &
0 \\
-2i \partial_{\bar{z}}& 0 & 0 & \Delta(r)\\
\bar{\Delta}(r) & 0 & 0 & 2i \partial_{z}\\
0 & \bar{\Delta}(r) & 2i \partial_{\bar{z}} & 0
\end{array} \right)
\end{displaymath}
$u_{a},v_{a}$ and $u_{b}, v_{b}$ denote the electronic (effective -
Dirac) variables of the two triangular sublattices, $a$ and $b$
respectively, of the honeycomb graphene lattice. For the usual
Kekule texture we get with $\Delta(r) = \Delta_{o}$. With no texture
we have two cones in the spectrum, $\epsilon_{\pm}(\vec{p}) = \pm
|\vec{p}|$, and with the Kekule texture mass gaps are opening :
$\epsilon_{\pm}(\vec{p}) = \pm \sqrt{|\vec{p}|^{2} +
|\Delta_{o}|^{2}} $ in the single particle spectra.
\subsection{Charge fractionalization}
Let's assume a vortex structure in the complex parameter
$\Delta(\vec{r})$ \cite{hou}:
\begin{equation}
\Delta(\vec{r}) = \Delta(r) \exp\{-i n \theta\}, \label{twist}
\end{equation}
where polar coordinates are used. We seek solutions for electronic
states of the dimerized graphene in the presence of this structure
with zero energy. The equations that follow from Eq.(\ref{ham}), in
the case of sublattice $a$ are
\begin{eqnarray}
\partial_{z}u + i \Delta(\vec{r}) v = 0 \nonumber \\
i \bar{\Delta}(\vec{r}) u - \partial_{\bar{z}}v = 0
\end{eqnarray}
and with the exchange $ z \leftrightarrow \bar{z} $ they are also
valid in the case of sublattice b.

In the polar coordinates we have
\begin{eqnarray}
\exp\{-i \theta\} (\partial_{r} - \frac{i}{r} \partial_{\theta})
u(\vec{r}) + i \exp\{-i n \theta\} \Delta(r) v(\vec{r}) = 0
\nonumber \\
i \exp\{i n \theta\} \bar{\Delta}(r) u(\vec{r}) -\exp\{i \theta\}
(\partial_{r} + \frac{i}{r} \partial_{\theta}) v(\vec{r}) = 0
\end{eqnarray}
In order to separate angular dependence we substitute $ u(\vec{r}) =
u_{o} \exp\{-i m \theta\} u(r) $ and $ v(\vec{r}) = v_{o} \exp\{-i l
\theta\} v(r) $ to have
\begin{eqnarray}
(\partial_{r} - \frac{m}{r}) u(r) + i \Delta(r) \frac{v_{o}}{u_{o}}
v(r) = 0 \nonumber \\
(\partial_{r} - \frac{l}{r}) v(r) + i \bar{\Delta}(r)
\frac{u_{o}}{v_{o}} u(r) = 0
\end{eqnarray}
if $l = n - 1 - m$. Further if we take $i \Delta(r)
\frac{v_{o}}{u_{o}} = f(r) \equiv |\Delta(r)|$ that fixes the ratio
$\frac{v_{o}}{u_{o}}$ the radial problem is reduced to solving
\begin{eqnarray}
(\partial_{r} - \frac{m}{r}) u(r)+ f(r) v(r) = 0 \nonumber
\\ (\partial_{r} - \frac{n - 1 - m}{r}) v(r) + f(r) u(r) = 0
\end{eqnarray}
There are two linearly independent solutions to the equations. The
behavior at large $r$ is, apart from powers of $r$, $ \exp\{\mp \mu
r\}$. Since the solution must be normalizable only one is
acceptable. At the origin, the asymptots that follow, with $f(r)
\sim f_{o} r^{|n|}$,
\begin{eqnarray}
u(r) = u_{1} r^{m} + u_{2} r^{|n|+ n - m} \nonumber \\
v(r) = v_{1} r^{|n|+ 1 + m} + v_{2} r^{n - 1 - m}
\end{eqnarray}
To have single-valued, non-singular solutions at the origin we have
to demand that $m$ is integer and
\begin{equation}
 n -1 \geq m \geq 0.
 \label{cond}
\end{equation}
The solutions that we get are similar but not the same as in Ref.
\onlinecite{jr}. There superconducting couplings in the Dirac
lagrangian induce different signs in the angular dependence $ u \sim
\exp\{-i m \theta\}, v \sim \exp\{i (n - 1 - m) \theta\}$ which then
guarantees one angular momentum eigenstate per any odd value of the
vorticity, $v = - n$. In our case only $v = -1$ vorticity solution
is an angular momentum eigenvalue: $ m = 0$. In this case the radial
 problem is simplified and the explicit solution is
 \begin{equation}
 \Psi(\vec{r}) = C \left [\begin{array}{c}
 0\\ i \exp\{i \alpha\}\\ 1 \\ 0 \end{array} \right]
 \exp\{- \int_{0}^{r} f(r^{'}) dr^{'}\}
 \label{zmsol1}
 \end{equation}
where $\alpha$ is a constant defined by $\Delta(r) = |\Delta(r)|
\exp\{i \alpha\} = f(r) \exp\{i \alpha\}$ and $C$ is a normalization
constant. If we do not demand that the solutions have to be
eigenstates of angular momentum the condition, Eq.(\ref{cond})
ensures that we have $n$ zero mode solutions in the case of the
vortex with vorticity $ v = - n$ ($n$ positive) \cite{hou}. Thus
negative vorticity vortex states exist only on sublattice $a$, and
very similar analysis shows that only positive vorticity vortex
states exist on sublattice $b$ in which case again there are as many
zero modes as the value of vorticity is \cite{hou}.

It can be argued \cite{hou} that the charge bound to a vortex of
vorticity $v = 1$ is $ - \frac{e}{2}$. We have to study the change
in the local density of states of the Dirac Hamiltonian \ref{ham} in
the presence of mass twist in Eq.(\ref{twist}). Because of the
sublattice symmetry \cite{hou} to any negative eigenstate of the
Dirac kernel $\Psi_{-\epsilon}(r)$ corresponds to a positive energy
state $\Psi_{\epsilon}(r)$, related to $\Psi_{-\epsilon}(r)$ by a
unitary transformation. Hence the local density of states
\begin{equation}
\nu(r, \epsilon) = \sum_{\epsilon^{'}} \Psi^{+}_{\epsilon^{'}}(r)
\Psi_{\epsilon^{'}}(r) \delta(\epsilon - \epsilon^{'})
\end{equation}
is symmetric with respect to zero energy. Demanding the conservation
of the total number of states after the inclusion of the mass twist,
we get
\begin{equation}
\int d\vec{r} \{2 \int_{- \infty}^{0^{-}} \delta \nu (\vec{r},
\epsilon) d\epsilon + | \Psi_{o}(\vec{r})|^{2}\} = 0,
\end{equation}
where $\Psi_{o}(\vec{r})$ stands for the single zero mode. Its
normalization to one leads to
\begin{equation}
\int d\vec{r}  \int_{- \infty}^{0^{-}} \delta \nu (\vec{r},
\epsilon) d\epsilon = - \frac{1}{2}
\end{equation}
so the net charge difference is $ - \frac{e}{2}$.
\subsection{Statistics}
The introduced system can be in short described as Dirac electrons
in the presence of a twisted mass \cite{sar}:
\begin{equation}\label{beginning_lagrangian}
{\cal L} = \bar{\Psi}(i \gamma_{\nu} \partial_{\nu} + \Delta \exp\{i
\gamma_{5} \phi\}) \Psi
\end{equation}
where $\gamma_{\nu}, \nu = 0,1,2$ are $4 \times 4$ Dirac matrices in
the Weyl representation. The problem may be reformulated by dividing
the vortex excitations into two groups, + and -, according to + and
- value of vorticity corresponding to singularities in $\phi_{+}$
and $\phi_{-}$ respecively, where $\phi = \phi_{+} + \phi_{-}$, and
introducing gauge fields \cite{comment},
\begin{eqnarray}
\label{defs} a_{\mu} = \frac{1}{2}(\partial_{\mu} \phi^{+} -
\partial_{\mu}
\phi^{-}) \nonumber \\
b_{\mu} = \frac{1}{2}(\partial_{\mu} \phi^{+} + \partial_{\mu}
\phi^{-})
\end{eqnarray}
We denote by $sgn(m_{3})$ \cite{sar} a quantity that at the singular
point of any vortex takes + or - depending whether the charge of the
vortex is of sublattice $a$ or $b$ kind \cite{comment2}. $sgn(m_{3})
\frac{\partial \times b}{\pi}$ represents vortex excitation electric
charge current:
\begin{equation}
\tilde{j}^{\mu} = \frac{sgn(m_{3})}{2 \pi} \epsilon^{\mu \nu
\lambda} \partial_{\mu} \partial_{\lambda} \phi.
\end{equation}
$sgn(m_{3})$ is necessary as we found out that vortex excitations
with both positive (+1) and negative (-1) vorticity that live on
sublattice $b$ and sublattice $a$ respectively, possess  the zero
mode and unoccupied by electron represent $- \frac{e}{2}$ absence of
charge.

$sgn(m_{3}) \frac{\partial \times a}{\pi}$ represents current of
axial charge (valley index) \cite{sar} associated with vortex
excitations (normalized by vortex axial charge) -
$\tilde{j}^{\mu}_{5}$. The sign of axial charge in the
density-current $\tilde{j}^{\mu}_{5}$ comes from $sgn(m_{3})$. This
comes from the correspondence - see Appendix, in the sign of the
sublattice density difference $ \bar{\Psi} \gamma_{3} \Psi$ (our
states are eigenstates of $\gamma_{0} \gamma_{3}$ \cite{jackpi}) and
the sign of the expectation value of $ \bar{\Psi} \gamma_{0}
\gamma_{5} \Psi$ - axial charge density. The expectation value of
the axial charge is $\pm \frac{1}{2}$ like of ordinary charge.

Then a simple statement follows for the topological part of the
effective action in the dual representation of the theory i.e. in
terms of vortices instead of Dirac particles (electrons). (In the
dual picture elementary ($2\pi$) fluxes of gauge fields are
particles and the gauge fields represent particle background.) The
form of the topological part of the action is

\begin{equation}\label{topological_part_of_the_action}
\frac{sgn(m_3)}{2 \pi} a (\partial \times b)- \frac{1}{2} a
\tilde{j}^{\mu}-\frac{1}{2} b\tilde{j}_5^{\mu}.
\end{equation}

We just encoded in Eq.(\ref{topological_part_of_the_action}) the
expressions of our currents found previously. There is overall
$\frac{1}{2}$ factor because of the value of the charge of the
vortices that must couple as $ - \frac{1}{2} A^{\mu}
\tilde{j}_{\mu}$ to the external field if we introduce it ($ a
\rightarrow a + A$).

If we introduce gauge fields
\begin{eqnarray}\label{R_L}
R=a+b, L=a-b
\end{eqnarray}
we arrive at the following form of the Lagrangian,
\begin{equation}
\frac{sgn(m_3)}{8 \pi}(R\partial R-L\partial
L)-R(\frac{\tilde{j}^{+}}{2})-L(\frac{\tilde{j}^{-}}{2}).
\end{equation}
where
\begin{equation}
\tilde{j}^{+} = \frac{\tilde{j}+\tilde{j}_{5}}{2}
\end{equation}
and
\begin{equation}
\tilde{j}^{-} = \frac{\tilde{j}-\tilde{j}_{5}}{2}, \label{end}
\end{equation}
the currents of the good quantum number - vorticity as opposed to
charge $\tilde{j}$ and axial $\tilde{j}_{5}$ currents. Considering
the Aharonov-Bohm phases for encircling quasiparticles around each
other we easily and clearly get that quasiparticles, $\tilde{j}^{+}$
and $\tilde{j}^{-}$, have semionic statistics among themselves and
their mutual statistics is trivial.

Therefore through a straightforward analysis of quasiparticles as
point particles i.e. singularities of phase - vortex solutions we
come
 to the conclusion that they obey semionic statistics and their
 theory is the doubled Chern-Simons i.e. $U(1)_{2} \times
 \overline{U(1)}_{2}$ - see Ref. \onlinecite{free}, as found in Ref. \onlinecite{sar}.
 On the other hand in Ref. \onlinecite{cha} the quasiparticles were
 viewed as extended objects - meron configurations of field
 $\vec{n}$, $\vec{n}^{2} = 1$, and the conclusion was that they
 possess
  quarton statistics. If we apply a simplification that
 vector $\vec{n}$ is always in the $x - y$ plane except at the
 center of excitation we will find following the arguments of
 Ref. \onlinecite{cha} that the excitation possess semionic statistics.
 Considering this we may ask ourselves what modifications of our
 approach are necessary to account for extended vortices.

The electric charge current $j^{\mu}$ and axial charge current
$j^{\mu}_{5}$ can be introduced in the topological part of the
action by simply taking $ j^{\mu} = \frac{1}{2} \tilde{j}^{\mu}$ and
$ j^{\mu}_{5} = \frac{1}{2} \tilde{j}^{\mu}_{5}$ because the point
charges carry half of the unit of electric and axial charge:
\begin{eqnarray}
& & \frac{sgn(m_3)}{2 \pi} a (\partial \times b)- a_{\mu}
j^{\mu}- b_{\mu} j_5^{\mu} \nonumber \\
&=&\frac{sgn(m_3)}{8 \pi}(R\partial R-L\partial L)-R j^{+}- L j^{-}.
\end{eqnarray}
 Now
we can see what principle can guide us to modify the theory. $j^{+}$
and $j^{-}$ should, in principle, correspond to the charged
fermions, electrons that may appear even in low-energy theory (they
should certainly exist in a complete theory). To have that we will
add appropriate Chern-Simons term as additional dynamics that is
allowed :
\begin{equation}
 \frac{sgn(m_3)}{2 \pi} a (\partial \times b) + \frac{sgn(m_3)}{2
\pi} a (\partial \times b)- a_{\mu} j^{\mu}- b_{\mu} j_5^{\mu}.
\label{new}
\end{equation}
Now fractionalized excitations, $ \tilde{j}^{\mu} = 2 j^{\mu}$ and $
\tilde{j}^{\mu}_{5} = 2 j^{\mu}_{5}$, have quarton statistics i.e.
\begin{equation}
\frac{sgn(m_3)}{4 \pi}(R\partial R-L\partial L)-\frac{1}{2} R
\tilde{j}^{+}- \frac{1}{2} L \tilde{j}^{-}.
\end{equation}
 This theory alone (as a topological one - so-called BF
Chern-Simons field theory and without $sgn(m_3)$ which can be
absorbed by simple redefinitions) was investigated in Ref.
~\onlinecite{hans} as the theory of 2d superconductors with vortices
and quasiparticles as excitations, and was proposed as the
description of the topological part of a phase for the QH
bilayer.~\cite{mvmzp} These considerations also imply that quartons
can be only found confined in pairs.

The beginning Lagrangian in (\ref{beginning_lagrangian}) can be
restated by a gauge transformation: $\Psi_{\pm}\rightarrow
e^{i\phi_{\pm}}\Psi_{\pm},$ where $\Psi_{\pm}$ are chiral components
of Dirac field, $\Psi_{\pm}=\frac{1}{2}(1\pm \gamma_5)\Psi$ as
\cite{sar}
\begin{equation}
\mathcal{L} = \overline{\Psi}(i\not\partial -\not a - \gamma_5 \not
b + \Delta)\Psi. \label{simple}
\end{equation}
After integrating out Dirac fermions, the total $\mathcal{L}$ can be
expressed also as \cite{sar}
\begin{eqnarray}
\mathcal{L}&=& -\frac{\pi}{12 \Delta} (\partial \times a)^{2} +
\frac{\Delta}{2 \pi} b^{2} + \frac{sgn(m_3)}{2 \pi} a (\partial
\times b) \nonumber \\
& & - \frac{1}{2} \tilde{j}^{\mu} a_{\mu} - \frac{1}{2}
\tilde{j}^{\mu}_{5} b_{\mu}.\label{sc}
\end{eqnarray}
When $\Delta$ or screening charge (in the Maxwell term) is large we
may expect that the point-like description (via semions) of vortices
is appropriate, but when $\Delta$ is small we are in a superfluid
phase where presumably quartons, but certainly some extended
objects, are confined and appear in pairs. This latter physics
remind us of the quantum Hall bilayer physics where merons in the
superfluid phase for the bilayer are extended objects, appear in
pairs, and have quarton statistics \cite{moo}. And indeed the action
(after integrating out fermions in the presence of an additional
field - a staggered chemical potential i.e. a third component of the
$\vec{n}$ vector) that was found in Ref. \onlinecite{cha}, is
similar to the effective action for the quantum Hall bilayer (as in
Ref. \onlinecite{moo}) where also
the fermionic current is equal to the $O(3)$ topological current
\cite{cha}. The theory in Eq.(\ref{new}) maybe a crude
oversimplification, but it tells us what is the feature of any phase
which includes both, fractionalized excitations (merons) and
fermions - the excitations must be bound in pairs (compare Ref.
\onlinecite{cha}). In other words when merons are deconfined, they
can be viewed as point-like objects with semionic statistics. That
was also a result of the numerical study in Ref. \onlinecite{apmvm}
of small (quantum dot) systems with spin in which deconfinement of
merons was proposed and mapped to a spinon (semion) gas of
Haldane-Shastry chain.

The conclusion is that the difference between Ref. \onlinecite{cha}
and Ref. \onlinecite{sar} in assigning the exchange statistics comes
as a difference in how one considers $\vec{n}$: (1) as a continuous
vector field as in the quantum Hall bilayer where quarton statistics
(meron - extended description) will survive even with no bias that
is, in Ref. \onlinecite{cha}, with no uniform $\mu_{s}$ - staggered
chemical potential but with adjustments of $\mu_{s}$ at places of
excitations, or (2) as a field that has to take values only in the
$x - y$ plane as in Ref. \onlinecite{sar} (and only uniform,
constant value of $\mu_{s}$ is allowed) with singular behavior on
vortices and hence semionic statistics. So the question is whether
we are promoting $\mu_{s}$ into a dynamical variable. Both
possibilities seem allowed but lead to different phases in general,
in which fractional objects have same charge but different
statistics, and are confined and deconfined respectively.

The derivation of the semionic statistics (Eqs. \ref{defs} -
\ref{topological_part_of_the_action}(\ref{end})) that we gave is
very general and still valid even with inclusion of a time reversal
breaking term - $\eta \bar{\Psi} \gamma_{5} \Psi$ that was included
and discussed in Ref. \onlinecite{cha}. The only assumption is the
value of the charge of the excitation for that case, which we can
safely take to be $\frac{1}{2}$ in accordance with the $\vec{n}$
formalism in Ref. \onlinecite{cha} (for $\mu_{s}$ = 0). A related
question or comment may be that the gauge transformation of Ref.
\onlinecite{sar} of ${\cal L}$ in Eq.(\ref{beginning_lagrangian})
will not lead to a simple transformed form with only fields: $\Psi,
a_{\mu},$ and $b_{\mu}$ (see Eq.(\ref{simple})) in this case. That
is true but it is very unlikely that (small) $\eta$ perturbation
will lead to an effective Lagrangian of $\Psi, a_{\mu}$, and
$b_{\mu}$ fields with a change of the coefficient of the minimal
coupling of field $a_{\mu}$ from one to two that is needed for
quarton statistics if we follow the same steps as in the derivation
of the semionic statistics. Therefore, the time reversal breaking
term alone can not lead to a statistical transmutation of semions
into quartons, although it seems its presence is the only way to
recover and demonstrate (Ref. \onlinecite{cha}) quarton statistics
in the $\vec{n}$ formalism ( when ($\Delta > |\eta|$)). The
inclusion of the uniform staggered chemical potential will change
the overall factor $\frac{1}{2}$ (the charge of the excitation) that
multiplies Eq.(\ref{topological_part_of_the_action}), but still the
statistical angle will be given by $ \frac{\theta}{\pi} = Q_{s}$
where $Q_{s}$ is the charge of the excitation ( and not by $
\frac{\theta}{\pi} = Q_{s}^{2}$).

\section{Fractionalization in bilayer graphene}

The question is what are physical systems that may support
fractionalization - obviously dimerized graphene is a hypothetical
system. The bosonic degrees of freedom that we need can come as a
result of electronic correlations, most notably excitonic and
superconducting. From these, only excitonic conserve charge and may
produce in their defects charged zero modes (as opposed to neutral
zero modes in superconductors) that we need to have
fractionalization.

In order to facilitate the discussion of the excitonic instability
and its zero modes in graphene bilayer, we will first discuss zero
modes in the case of $p$-wave superconducting and excitonic system.

(1) The effective BCS Hamiltonian for the quasiparticles is
\begin{equation}
H=\sum_{k}\xi_k c_k^{+}c_k +
\frac{1}{2}(\Delta_k^{*}c_{-k}c_k+\Delta_{k}c_{k}^{+}c_{-k}^{+}).
\end{equation}

Introducing the Bogoliubov transformation:
\begin{equation}
\alpha_{k}=u_k c_k - v_k c_{-k}^{+}
\end{equation}

that should diagonalize the Hamiltonian into $H=\sum E_k
\alpha_k^{+}\alpha_{k}+const$, implies Bogoliubov-de Gennes
equations

\begin{eqnarray}
E_k u_k = \xi_k u_k - \Delta_k^* v_k, \\
E_k v_k = -\xi_k v_k -\Delta_k u_k.
\end{eqnarray}

If $\Delta_k=\Delta(k_x-ik_y)$ and $\xi_k \approx -\mu (\mu>0)$, in
the long-distance approximation the equations for zero mode(s)
become:

\begin{eqnarray}\label{zero_modes}
-\mu u - \Delta(-i)\partial_{\overline{z}}v=0,\\
\mu v - (-i)\Delta\partial_{z}u=0. \label{zero_modes_end}
\end{eqnarray}

We may ask for zero modes that exist in vortex solutions for which
we demand $v(\theta+2\pi)=-v(\theta)$ and
$u(\theta+2\pi)=-u(\theta)$ in polar coordinates~\cite{read}. By
solving (\ref{zero_modes}) with $\Delta=const$ we are neglecting the
short (small radial) distance details of the solution. We seek the
solution in the following form

\begin{equation}\label{solution_form}
u=\frac{u(r)}{\overline{z}^l}, v=\frac{v(r)}{z^k}
\end{equation}

and the equations that we get are

\begin{eqnarray}
-\mu
\frac{u(r)}{\overline{z}^l}+i\Delta\frac{1}{z^k}e^{i\theta}\partial_r
v(r)=0,\\
\mu\frac{v(r)}{z^k}+i\Delta\frac{1}{\overline{z}^l}e^{-i\theta}\partial_r
v(r)=0.
\end{eqnarray}

To separate angular and radial dependence we must have
$k=l=\frac{1}{2}$ and the equations are reduced to

\begin{eqnarray}
-\mu u(r)+i\Delta \partial_r v(r) =0,\\
\mu v(r)+i\Delta \partial_r u(r)=0.
\end{eqnarray}

If $u(r)=u_0 f(r)$ and $v(r)=v_0 f(r)$, the equations reduce to a
single one,

\begin{equation}
\mu f(r)+\Delta\partial_r f(r)=0,
\end{equation}

if $\frac{u_0}{v_0}=-i.$ Therefore our solutions can be cast in the
following form:

\begin{equation}
u=ie^{\frac{\pi}{4}}\frac{f(r)}{\sqrt{\overline{z}}},
v=e^{i\frac{\pi}{4}}\frac{f(r)}{\sqrt{z}},
\end{equation}

where $f(r)$ is of the simple radial dependence $\sim
e^{-\frac{\mu}{\Delta}r}$ for $\mu=const.$

The usual approach~\cite{stone,tew} is to model order parameter with
the vortex singularity i.e. to take, instead of $\Delta_k$,
$e^{i\frac{\theta}{2}}\Delta_k e^{i\frac{\theta}{2}}$ in our case.
This symmetrized with respect to phase expression is used to ensure
the antisymmetry of the order parameter i.e. that the term $\int
d\vec{r} \Psi^{+}(\vec{r})\Delta_k \Psi^{\dagger}(\vec{r})$ in the
Bogoliubov Hamiltonian is well-defined and consistent with the
anticommutativity of fermi operators when phase is
coordinate-dependent.

From the approach we used in getting the zero mode we can turn to
the usual approach by the simple phase transformation $u\rightarrow
e^{-i\frac{\theta}{2}}u, v\rightarrow e^{i\frac{\theta}{2}}v,$ so
that at the end our zero mode solution has the components:

\begin{equation}
u=i e^{i\frac{\pi}{4}}\frac{f(r)}{\sqrt{r}},
v=e^{i\frac{\pi}{4}}\frac{f(r)}{\sqrt{r}}
\end{equation}

Now the quasiparticle operator for the zero energy state can be
written as

\begin{equation}
\gamma_0^{+}=\int d^2\vec{r} ( u(\vec{r})c^{+}(\vec{r}) +
v(\vec{r})c(\vec{r}) )
\end{equation}

and immediately we can conclude that $\gamma_0^+=\gamma_0$ for our
solution i.e. it represents neutral Majorana mode.

We should notice that with respect to the Dirac problem in the
dimerized graphene here, in the latter approach, momentum operators
are together, in the same term, with order parameter and phase
singularity in the Hamiltonian.

On the other hand, in the case of the excitonic problem which may be
related and become a physical realization of dimerized graphene
lattice~\cite{hou}, we have a different basic Hamiltonian:

\begin{equation}\label{different_basic_hamiltonian}
H=\sum_{k}E_k
(\beta_k^+\beta_k-\gamma_k^+\gamma_k)-\sum_{k}(\Delta_k
\beta_k^+\gamma_k+\Delta_k^* \gamma_k^+\beta_k )
\end{equation}

Introducing

\begin{eqnarray}
B_k = u_k \beta_k - v_k \gamma_k, \\
C_k = v_k \beta_k+u_k\gamma_k
\end{eqnarray}

that should diagonalize the Hamiltonian into
$H=\sum_{k}\epsilon_k(B_k^+B_k-C_k^+C_k),$ we get

\begin{eqnarray}\label{epsilon_u_v}
\epsilon_k u_k = E_k u_k + v_k \Delta_k^*,\\
\epsilon_k v_k = E_k v_k - u_k \Delta_k.
\end{eqnarray}

For the zero modes ($\epsilon_k=0$) simple redefinition $v_k
\rightarrow -v_k$ transforms the equations into the same as for the
superconducting problem.

We will assume $\Delta_k=\Delta(k_x-ik_y)$ and
$E_k=\epsilon=const>0$. Then for the zero modes we have

\begin{eqnarray}
\epsilon u +\Delta(-i)\partial_{\overline{z}}v=0,\\
\epsilon v-\Delta(-i)\partial_{z}u=0.
\end{eqnarray}

The equations are the same as in Eq.(\ref{zero_modes}) and
Eq.(\ref{zero_modes_end}). Again, we may ask for the zero modes that
exist in vortex solutions for which we demand
$v(\theta+2\pi)=-v(\theta)$ and $u(\theta+2\pi)=-u(\theta)$ in polar
coordinates and the answer would be the same. But if we want to stay
in the language of the order parameter, we may model it, in the
presence of a vortex, as $\Delta_k e^{i\theta}$, with no
symmetrization as was necessary in the superconducting problem. In
this case the solutions become

\begin{equation}
u=ie^{i\frac{\pi}{4}}e^{i\frac{\theta}{2}}\frac{f(r)}{\sqrt{r}},
v=e^{i\frac{\pi}{4}}e^{i\frac{\theta}{2}}\frac{f(r)}{\sqrt{r}}
\end{equation}

which are not satisfactory for the electronic wave functions because
they are not single-valued while the singularity is borne by the
order parameter. If we choose in the order parameter
$e^{-i\frac{\theta}{2}}\Delta_k e^{i\frac{\theta}{2}}$, our
solutions are represented by

\begin{equation}
u=ie^{i\frac{\pi}{4}}e^{i\theta}\frac{f(r)}{\sqrt{r}},
v=e^{i\frac{\pi}{4}}\frac{f(r)}{\sqrt{r}}
\end{equation}

and this, although not an angular momentum eigenstate, is the
solution that can describe a charged mode because $u^* \neq v$ as a
crucial difference with respect to the superconducting case. The
excitonic Bogoliubov transformation mixes the same kind of charged
operators, leading to charged zero modes as in the dimerized
graphene problem. Although the precise form of the equations for the
zero modes is not the same in the two cases (excitonic $p$-wave
system and dimerized graphene), they are very similar in the
long-distance regime and should lead to the same conclusion about
the charge fractionalization.

(2) For the case of bilayer graphene excitonic instability described
in Ref.~\onlinecite{han}, we have in the limit of small interlayer
hopping

\begin{equation}
E_k \approx V\left[ 1-\frac{1}{2}(\frac{\epsilon_k}{t_{\perp}})^2
\right]
\end{equation}

for the energy in Eqs.(\ref{epsilon_u_v}). Here $t_{\perp}$ is the
interlayer hopping parameter, $\pm V$ is the bias that causes excess
electrons and holes in the lower and upper graphene layers,
respectively, and $\epsilon_k$ is the bare kinetic energy for which
$\epsilon_k \sim k.$ The gap function is of the following form, near
two nodal points~\cite{han},

\begin{equation}
\Delta_k = i(k_x\mp ik_y)|k|\Delta,
\end{equation}

where $\Delta$ is a positive constant. We will write $E_k$ as
$E_k=V-\delta k^2$ with also $\delta$ being a positive constant.
Then the equations (\ref{epsilon_u_v}) with $\epsilon_k=0$ become

\begin{eqnarray}
(V+\delta\frac{\partial}{\partial z}\frac{\partial}{\partial
\overline{z}})u + \Delta(-i)(-i)\sqrt{\frac{\partial}{\partial
z}\frac{\partial}{\partial\overline{z}}}
(-i)\partial_{\overline{z}}v=0,\\
(V+\delta\frac{\partial}{\partial z}\frac{\partial}{\partial
\overline{z}})v - \Delta(i)(i)\sqrt{\frac{\partial}{\partial
z}\frac{\partial}{\partial\overline{z}}} (-i)\partial_{z}u=0.
\end{eqnarray}

To consider the vortex solution we have to set
$e^{-i\frac{\theta}{2}}\Delta_k e^{i\frac{\theta}{2}}$ instead of
$\Delta_k$, which if we want to absorb the phase into $v$ and $u$
leads to slightly different equations due to the presence of the
$k^2$ terms that we also must keep in this long-distance analysis.
Because

\begin{eqnarray}
\nonumber e^{i\frac{\theta}{2}} \frac{\partial}{\partial z}
\frac{\partial}{\partial \overline{z}} e^{-i\frac{\theta}{2}}=
e^{i\frac{\theta}{2}} \frac{\partial}{\partial z}
e^{-i\frac{\theta}{2}}e^{i\frac{\theta}{2}}\frac{\partial}{\partial
\overline{z}} e^{-i\frac{\theta}{2}} \\
= (\frac{\partial}{\partial
z}-\frac{1}{2z})(\frac{\partial}{\partial\overline{z}}+\frac{1}{2\overline{z}}),
\end{eqnarray}

the equations for the vortex solution become

\begin{eqnarray}
\nonumber \left[ V+\delta(\frac{\partial}{\partial z}
-\frac{1}{2z})(
\frac{\partial}{\partial\overline{z}}+\frac{1}{2\overline{z}})\right]
u \\ + \Delta(+i) \sqrt{\frac{\partial}{\partial
z}\frac{\partial}{\partial \overline{z}}}
\partial_{\overline{z}}v=0, \label{vortex_sol1} \\
\nonumber \left[ V+\delta(\frac{\partial}{\partial z}
-\frac{1}{2z})(
\frac{\partial}{\partial\overline{z}}+\frac{1}{2\overline{z}})\right]
v \\ + \Delta(-i) \sqrt{\frac{\partial}{\partial
z}\frac{\partial}{\partial \overline{z}}}
\partial_{z}u=0, \label{vortex_sol2}
\end{eqnarray}

with the requirement $u(\theta+2\pi)=-u(\theta)$ and
$v(\theta+2\pi)=-v(\theta)$. If we seek solutions in the following
form (see Eq.(\ref{solution_form}))

\begin{equation}
u=\frac{u(r)}{\overline{z}^l}, v=\frac{v(r)}{z^k}
\end{equation}

we face the problem of properly performing the operations with the
square root operator, $\sqrt{\frac{\partial}{\partial
z}\frac{\partial}{\partial \overline{z}}}$. Its action we will
define on the space of monomials in $z$ and $\overline{z}$, for
which we have

\begin{equation}
\sqrt{z\frac{\partial}{\partial
z}}\sqrt{\overline{z}\frac{\partial}{\partial \overline{z}}} z^n
\overline{z}^m =\sqrt{n}\sqrt{m}z^n \overline{z}^m.
\end{equation}

We will assume the asymptotic behavior of $u(r)$ and $v(r)$ as
$u(r)\sim e^{-\lambda r}$ and $v(r)\sim e^{-\lambda r}$, and justify
it in the end. Then, for example, in the case of $v(r)$ we have

\begin{eqnarray}
\nonumber \sqrt{\frac{\partial}{\partial z}\frac{\partial}{\partial
\overline{z}}} \frac{1}{z^k}e^{i\theta}\partial_r v(r) = -\lambda
v_0 \sqrt{\frac{\partial}{\partial z}\frac{\partial}{\partial
\overline{z}}} \{ \frac{1}{z^k}\frac{z^{1/2}}{\overline{z}^{1/2}}
\\
\nonumber \times \sum_{n=0}^{\infty}\frac{1}{n!}
(-\lambda)^n z^{n/2}\overline{z}^{n/2}\} \\
=-\lambda v_0 \frac{1}{\sqrt{z\overline{z}}}
\sqrt{z\frac{\partial}{\partial z}}
\sqrt{\overline{z}\frac{\partial}{\partial \overline{z}}}
\\
\nonumber \times \sum_{n=0}^{\infty}\frac{1}{n!}(-\lambda)^n
z^{\frac{n+1}{2}-k}\overline{z}^{\frac{n-1}{2}}\\
=-\lambda v_0
\frac{1}{\sqrt{z\overline{z}}}\sum_{n=0}^{\infty}\frac{1}{n!}(-\lambda)^n
\\
\times
\sqrt{\frac{n+1}{2}-k}\sqrt{\frac{n-1}{2}}z^{\frac{n+1}{2}-k}\overline{z}^{\frac{n-1}{2}}.
\end{eqnarray}

Therefore the Eq.(\ref{vortex_sol1}) becomes

\begin{eqnarray}
\nonumber \left[ V+\delta(\frac{\partial}{\partial z}
-\frac{1}{2z})(
\frac{\partial}{\partial\overline{z}}+\frac{1}{2\overline{z}})\right]
u(r) \\
\nonumber + \Delta i (-\lambda) v_0
\frac{1}{\sqrt{z\overline{z}}}\sum_{n=0}^{\infty}\frac{1}{n!}(-\lambda)^n
\times \\
\sqrt{\frac{n+1}{2}-k}\sqrt{\frac{n-1}{2}}z^{\frac{n+1}{2}-k}\overline{z}^{\frac{n-1}{2}+l}
= 0
\end{eqnarray}

To factor the angular dependence we choose $k=l=\frac{1}{2}$ and in
the long distance approximation, for which we take also
$\sqrt{n(n-1)}\approx n$ in the series expansion, we have

\begin{eqnarray}
\left[ V+\delta \partial_r^2 \right] u + i \Delta (-\lambda)^2 v(r)
= 0, \\
\left[ V+\delta \partial_r^2 \right] v -i \Delta (-\lambda)^2 u(r)
=0.
\end{eqnarray}

This means

\begin{eqnarray}
u_0 \left[ V+\delta \lambda^2 \right] + i\Delta \lambda^2 v_0 =0, \\
v_0 \left[ V+\delta \lambda^2 \right] - i\Delta \lambda^2 u_0 =0 \\
\end{eqnarray}

i.e. $\lambda_1^2=\frac{V}{-(\delta+\Delta)}$ or
$\lambda_2^2=\frac{V}{\Delta-\delta}$. Both, $\lambda_1=\pm
i\sqrt{\frac{V}{\delta+\Delta}}$, represent delocalized zero modes.
In the case of $\lambda_2$ that would be if $\delta>\Delta$,
otherwise we have only one physical localized state with
$e^{-\lambda_2 r }, \lambda_2=\sqrt{\frac{V}{\Delta-\delta}}$, decay
function. Comparing with solutions in Ref.\onlinecite{han} for
$\Delta$ for fixed parameters for the graphene bilayer, we find that
for large enough $V$, bias parameter, we can have the bound
solution. Then the three zero modes  carry fractional charge,
$\frac{3}{2}$, that we get applying the same arguments that were
given in Ref.\onlinecite{hou} for the dimerized graphene. Essential
in these arguments is that Dirac kernel has the correspondence of
positive and negative energy solutions. In this case we also have
the Dirac structure of the problem, with the nodal points, and the
same correspondence.

In this Hartree-Fock~\cite{han} treatment of the excitonic problem
we get three zero modes (for large enough $V$) and the question
comes whether they will stay with further inclusion of interactions.
In the case of three zero modes, the charge that they carry is
mostly smeared out through the system. If the interactions were able
to split the two delocalized zero modes, we could have localized
charge $\frac{1}{2}$ vortex excitation similarly to the case of
dimerized graphene.~\cite{hou} The conclusions about the topological
terms and fractional statistics that we reached in the case of
dimerized graphene will still hold. The interaction term that we
have in mind would be $\sim V \rho_{k_{o}}\rho_{-k_{o}}$ where
$k_{o} = |\lambda_{1}|$ and an excitonic coupling between the mode
$k_{o}$ and $-k_{o}$ would produce necessary splitting. Otherwise
(with no splitting) the situation is less clear but if we assume
that the localized mode describes a missing of $\frac{e}{2}$ charge
the delocalized modes will describe additional degrees of freedom
(that may be occupied or unoccupied) that may further decrease the
value of the statistical angle of excitations or equivalently
increase the coupling of the doubled Abelian Chern-Simons term.

The recent work, Ref. \onlinecite{baba}, that appeared while we were
finishing the writing, concerns fractionalization in excitonic
bilayer graphene that is not naturally (Bernal) stacked but consists
of two parallel layers at some larger distance that leads to even
number of zero modes due to the valley degeneracy and, therefore, no
charge fractionalization. Still we can not rule out, on the basis of
the long distance analysis of vortex solutions in the excitonic
condensate that we presented, the same doubling in our case. This
important question can be resolved only by a detailed, numerical
analysis of the bilayer graphene.

\section{Acknowledgment} The author thanks Chang-Yu Hou for correspondence.
The work was supported by Grant No. 141035
of the  Ministry of Science of the Republic of Serbia.

\appendix*
\section{}
In the second quantized formalism the zero mode solution in
Eq.(\ref{zmsol1}), of definite vorticity $v = -1$, contributes to
the expansion of the Dirac field as a term equal to
 \begin{equation}
 \Psi_{o}(\vec{r}) =  \left\{ \left[\begin{array}{c}
 0\\ i \exp\{i \alpha\}\\ 0 \\ 0 \end{array}\right] c_{0} + \left [\begin{array}{c}
 0\\ 0 \\ 1 \\ 0 \end{array}\right] c_{0}^{+} \right\} C
 \exp\{- |\Delta| r \},
 \label{zmsol2}
 \end{equation}
 where we simplified the decay function by taking $|\Delta(r)| =
 |\Delta| = const$, and again the $C$ is the normalization constant.
 Notice the absence of two different operators (one for particle,
 the other for hole) that we would have for a non-zero energy level.
 For the sake of the argument, going in reverse, we can fix the
 normalization constant by demanding that the charge associated with
 the zero mode is $ (-) \frac{1}{2}$ i.e.
 \begin{equation}
 \int d^{2}\vec{r} < \Psi_{o}^{+}(\vec{r})\Psi_{o}(\vec{r})> = -
 \frac{1}{2}.
 \end{equation}
That would imply
\begin{equation}
 \int d^{2}\vec{r} < \Psi_{o}^{+}(\vec{r})\gamma^{5}\Psi_{o}(\vec{r})> =
 \frac{1}{2},
 \end{equation}
 for the expectation value of the axial charge, where we took
\begin{displaymath}
\gamma_{5} = \left[\begin{array}{cccc} 1&0&0&0\\
0&1&0&0\\0&0&-1&0\\0&0&0&-1 \end{array}\right]
\end{displaymath}
in the Weyl representation. On the other hand our state is an
eigenstate of the sublattice charge difference operator:
\begin{displaymath}
R = \alpha_{3} = \gamma_{0} \gamma_{3} =
\end{displaymath}
\begin{displaymath}
 \left[\begin{array}{cccc} 0&0&1&0\\
0&0&0&1\\1&0&0&0\\0&1&0&0 \end{array}\right]\left[\begin{array}{cccc} 0&0&-1&0\\
0&0&0&1\\1&0&0&0\\0&-1&0&0 \end{array}\right] =\left[\begin{array}{cccc} 1&0&0&0\\
0&-1&0&0\\0&0&-1&0\\0&0&0&1 \end{array}\right]
\end{displaymath}
so that
\begin{equation}
 \int d^{2}\vec{r} < \Psi_{o}^{+}(\vec{r})\alpha_{3}\Psi_{o}(\vec{r})> =
 \frac{1}{2},
 \end{equation}
 of the same sign as the expectation value of the axial charge. Both
 signs would reverse if the solution were with vorticity $v = 1$, on
 sublattice $b$, although the sign of the electric charge would
 remain the same.


\begin{references}
\bibitem{hou} C.-Y. Hou, C. Chamon, and C. Mudry, Phys. Rev. Lett. {\bf 98}, 186809
(2007).
\bibitem{sar} B. Seradjeh, C. Weeks, and M. Franz, Phys. Rev. B {\bf 77}, 033104 (2008);
B. Seradjeh and M. Franz, arXiv:0709.4258
\bibitem{jackpi} R. Jackiw and S.-Y. Pi, Phys. Rev. Lett. {\bf 98}, 266402 (2007).
\bibitem{cha} C. Chamon, C.-Y. Hou, R. Jackiw, C. Mudry, S.-Y. Pi,
and A.P. Schnyder, Phys. Rev. Lett. {\bf 100}, 110405 (2008).
\bibitem{su} W.P. Su, J.R. Schrieffer, and A.J. Heeger, Phys. Rev.
Lett. {\bf 42}, 1698 (1979).
\bibitem{comment} With respect to the Ref. \onlinecite{sar}  we fix
 that partitioning is into the different
vorticity groups. The vorticity is a good number and allows this
partitioning.
\bibitem{comment2} $sgn(m_{3})$ is another good quantum number, sign
of the charge carried by the zero mode (occupied or unoccupied).
There are four kinds of vortices depending on the $sgn(m_{3})$ and
vorticity.
\bibitem{han} R. Dillenschneider and Jung Hoon Han, arXiv:0709.1230
\bibitem{jr} R. Jackiw and P. Rossi, Nucl. Phys. B 190, 681 (1981).
\bibitem{free} M. Freedman, C. Nayak, K. Shtengel, K. Walker, and Z. Wang, Ann. Phys.
{\bf 310}, 428 (2004).
\bibitem{hans} T.H. Hansson, V. Oganesyan, and S.L. Sondhi, Ann.
Phys. {\bf 313}, 497 (2004).
\bibitem{mvmzp} M.V. Milovanovi\'{c} and Z. Papi\'{c}, arXiv:0710.0478
\bibitem{moo} K. Moon,
H. Mori, Kun Yang, S.M. Girvin, A.H. MacDonald, L. Zheng, D.
Yoshioka, and S.-C. Zhang, Phys. Rev. B {\bf 51}, 5138 (1995).
\bibitem{apmvm} A. Petkovi\'{c} and M.V. Milovanovi\'{c}, Phys. Rev.
Lett. {\bf 98}, 066808  (2007).
\bibitem{read} N. Read and D. Green, Phys. Rev. B  {\bf 61}, 10267
(2000).
\bibitem{stone} M. Stone and R. Roy, Phys. Rev. B   {\bf 69}, 184511 (2004); M. Stone
and S.-B. Chung, Phys. Rev. B  {\bf 73}, 014505 (2006).
\bibitem{tew} S. Tewari, S. Das Sarma, C. Nayak, C. Zhang, and P. Zoller, Phys. Rev. Lett. {\bf
98}, 010506 (2007).
\bibitem{baba} B. Seradjeh, H. Weber, and M. Franz, arXiv:0806.0849
\end{references}
\end{document}